\documentclass[9pt,conference]{IEEEtran}
\usepackage{dcase2025}

\usepackage{bm} 

\usepackage{dcase2025,amsmath,graphicx,url,times,booktabs, tabularx, tcolorbox}
\definecolor{moderngreen}{RGB}{0,100,0}

\title{Audio Question Answering with GRPO-Based Fine-Tuning and Calibrated Segment-Level Predictions}

\name{
Marcel Gibier, 
Nolwenn Celton,
Raphaël Duroselle, 
Pierre Serrano,
Olivier Boeffard,
Jean-François Bonastre
}

\address{
Inria Paris, LR2 \\
 Paris, France \\
\texttt{prenom.nom@inria.fr}
}

\usepackage[justification=centering]{caption}

\begin{document}

\maketitle

\begin{abstract}
In this report, we describe our submission to Track 5 of the DCASE 2025 Challenge for the task of Audio Question Answering (AQA). Our system leverages the SSL backbone BEATs to extract frame-level audio features, which are then processed by a classification head to generate segment-level predictions of acoustic events, following the Audioset ontology. These segment-level predictions are subsequently calibrated before producing event-level predictions. Finally, these predictions are incorporated into a structured prompt, along with the question and candidate answers.
This prompt is then fed to a fine-tuned version of Qwen2.5-7B-Instruct, trained using the GRPO algorithm with a simple reward function. Our method achieves an accuracy of 62.6 \% on the development set, demonstrating the effectiveness of combining acoustic event reasoning with instruction-tuned large language models for AQA.
\end{abstract}

\begin{IEEEkeywords}
Audio question answering, Sound event detection, Classification, Calibration, GRPO
\end{IEEEkeywords}

\section{Introduction}
\label{sec:intro}
The task of Audio Question Answering (AQA) involves providing an answer to a question about an audio clip. It has broad applications in human-computer interaction, accessibility, and multimedia retrieval. However, audio understanding remains challenging due to its temporal nature, variability, and the difficulty of aligning acoustic events with linguistic reasoning.
\vspace{0.3em}

High-performing solutions for AQA typically involve multimodal large language models (LLMs), and such models have been proposed as baselines: Qwen2-Audio-7B \cite{Chu2024}, AudioFlamingo2 \cite{ghosh2025}, and Gemini-2.0-Flash. These models typically extract audio representations from each frame using an encoder such as an Audio Spectrogram Transformer \cite{ast}. The representations are then projected into an embedding space matching the dimensionality of the language model’s textual inputs, which can be either shared or separate. Audio can be provided either as learned queries through a Q-Former\cite{qformer}, as a prefix to the text tokens, or via a cross-attention mechanism within the language model \cite{flamingo}, which is trained using causal language modeling to answer questions about the audio using datasets such as OpenAQA~\cite{ltu}. If the language model’s output does not exactly match one of expected answers, techniques such as regular expressions or similarity scoring with Sentence-BERT can be used to evaluate the response.
\vspace{0.3em}

Despite their effectiveness, these models rely on direct reasoning over high-dimensional audio features, which can limit interpretability and generalization. In this work, we propose an alternative approach based on symbolic reasoning. Instead of analyzing raw audio, we first extract a timestamped list of detected acoustic events and convert it into a textual description. This structured summary is then used by a language model to answer the question.
\vspace{0.3em}

This approach introduces two main challenges: (i) the language model must be capable of performing robust linguistic reasoning from structured event-based descriptions, and (ii) the reliability of the final prediction depends on the calibration of the confidence scores output by the acoustic event detection (SED) model. In other words, accurate reasoning depends on accurate and well-calibrated detection.
\vspace{0.3em}

We present two contributions in this work: (i) we incorporate a calibration method for segment-level sound event predictions to improve the reliability of the symbolic input to the language model; (ii) we investigate the benefits of fine-tuning using the Group Relative Policy Optimization (GRPO) algorithm, a variant of PPO, as an alternative to conventional Supervised Fine-Tuning (SFT) for the AQA task.
\begin{figure}[t]
  \centering
  \centerline{\includegraphics[width=\columnwidth]{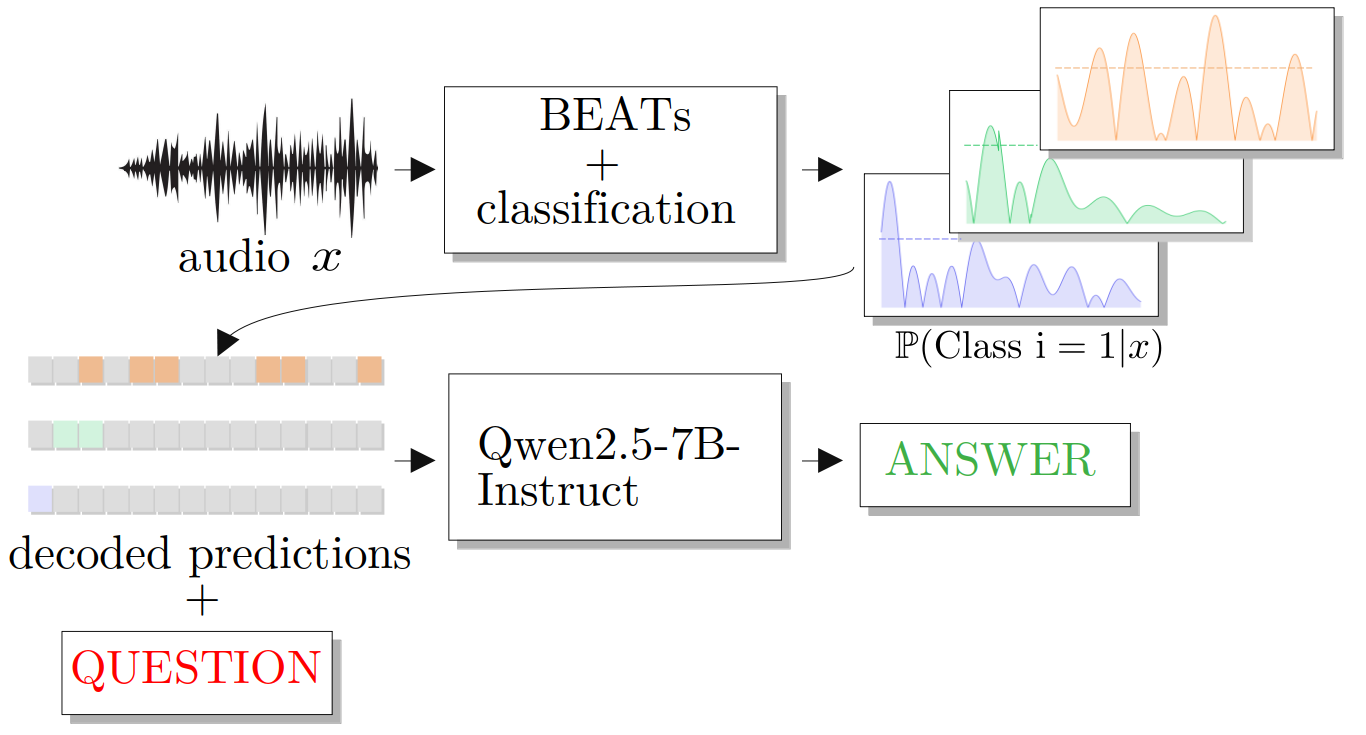}}
  \centering
  \caption{Our audio question-answering system is based on two main steps: (i) we decode the class presence probabilities output by the sound event detection model to obtain a time-localized, textual sound description of the audio; (ii) this text, along with the question, is then provided to a language model, allowing it to answer without directly analyzing the audio.}
  \label{fig:inference}
\end{figure}
\vspace{0.3em}

Our submitted system achieved an accuracy of 62.6\% on the development set, significantly outperforming the best baseline score of 52.5\% obtained by Gemini-2.0-Flash.

\section{Method}
\subsection{Calibration of segment-level sound event predictions}
Our system relies on a sound event detection (SED) system that outputs event-level predictions constituted of an event label, and start and end timestamps. We use the pretrained audio transformers proposed by \cite{schmid2025effective}. The SED system is constituted of several components: an audio encoder, a classification head at the segment-level that outputs posterior probabilities for each event class, and a post-processing algorithm that produces event-level predictions (median filtering and threshold on segment-level posterior probabilities).
\vspace{0.3em}

We observed that the values of the posterior probabilities on the DCASE corpus did not seem to reflect the relative proportion of target and non target segments. This property is called miscalibration.  It can lead to suboptimal event-level predictions because the same threshold applied to different classes can correspond to different compromises between false alarm and false rejection.\\

\subsubsection{Logistic regression}
Rather than calibrating the posterior probabilities directly, which are often biased by class frequency imbalances in the evaluation data, we chose to calibrate the likelihood ratios (LR). These scores provide a measure that is independent of prior distribution assumptions.
\begin{align}
\text{LR}(x) &=  \left( 
\frac{P(x \mid y = 1)}{P(x \mid y = 0)} 
\right)
\label{eq:llr}
\end{align}

To this end, we trained a separate logistic regression model for each class, aimed at transforming the raw LRs produced by the BEATs model into calibrated scores. We used the implementation of \cite{alumae2019taltech}. The reduction of the calibration error can be evaluated by comparing the values of a proper scoring rule such as $C_{llr}$ before and after calibration \cite{ferrer2024evaluating}.
 
 \begin{align}
C_{llr} = \frac{1}{2} \bigg( 
    & \frac{1}{N_{y=1}} \sum_{i} \log_2 \left(1 + \frac{1}{\text{LR}_{(y=1),i}} \right) \nonumber \\
  + & \frac{1}{N_{y=0}} \sum_{j} \log_2 \left(1 + \text{LR}_{(y=0),j} \right)
\bigg)
\label{eq:cllr}
\end{align}
with $\{y=1\}$ and $\{y=0\}$ indicating the segments with and without the target event.
\\
\subsubsection{Priors adjustment}
\label{sec:priors_calibration}
We estimated the class prior probabilities $P(y=1)$ using the class distribution observed in the training subset. This allowed us to convert the calibrated LLRs back into posterior probabilities, which were then evaluated with reliability curves. 

\begin{align}
P(y=1|x)= \frac{LR \times P(y=1)}{LR \times P(y=1) + (1 - P(y=1))}
\label{eq:logits_to_preds}
\end{align}
\vspace{0.1em}
\subsection{Using GRPO for Question Answering}
Reinforcement Learning from Human Feedback (RLHF) is a fine-tuning method used to align a language model's responses with human preferences. This approach is based on reinforcement learning, where the goal is to optimize a policy, that is, the generative model itself, by maximizing a reward function. To guide this optimization, a reward model trained on human feedback is used to evaluate the quality of the model’s responses. Traditionally, a reinforcement learning architecture also includes a critic, a separate neural network responsible for estimating the expected value of the actions taken by the policy. The critic helps stabilize training by providing a more accurate estimate of the advantage, which represents the difference between the received reward and the expected reward.
\vspace{0.3em}

We adopt Group Relative Policy Optimization (GRPO) for fine-tuning via RLHF \cite{deepseek}. In contrast to other reinforcement learning methods, GRPO removes the need for a separate critic model by using the average reward across multiple generations per prompt from the policy itself to establish a baseline for advantage estimation. This choice is also motivated by the fact that RLHF, particularly using GRPO, significantly outperforms supervised fine-tuning on downstream tasks with limited data \cite{grpoaqa, r1zero}.
\vspace{0.3em}

For each question $q$, in the format described in Box \ref{box:sap}, the model produces a set $G = \{ o_i \}$ of candidate answer tokens, where each $o_i \in \{ \text{A}, \text{B}, \text{C}, \text{D} \}$ corresponds to one of the multiple-choice options. We denote $\pi_{\theta}(o_i \mid q) = P(\text{token} = o_i \mid q)$ as the probability distribution over the elements of $G$ where $\theta$ represents the model's parameters.
\vspace{0.3em}

We define a simple reward function that assigns a value of 1 if the output $o_i$ is the correct answer, and 0 otherwise. Our objective is to assess whether the response $o_i$ outperforms the group’s average response. To encourage the model to increase the probability $\pi_\theta(o_i)$, the advantage function is defined as follows:
\begin{align}
\hat{A_i} = \frac{R_i - \bar{R}}{\sigma_R}
\end{align}
where $R_i$ is the reward for the $i$-th response, $\bar{R}$ is the mean reward across the group, and $\sigma_R$ is the corresponding standard deviation.

In the end, the goal is to increase the contribution of responses with high advantage, so we aim to maximize the following quantity, which is a bit simpler than the original DeepSeek-R1  method:
\begin{align}
\mathcal{J}_{\text{GRPO}}(\theta)
= &\ \mathbb{E}_{q \sim P(Q)}\left[
\begin{aligned}
&\frac{1}{|G|}\sum_{i=1}^{|G|}
\min\!\bigl(r_i \,\hat A_i,\ \mathrm{clip}(r_i,1-\epsilon,1+\epsilon)\,\hat A_i\bigr)
\end{aligned}
\right] \nonumber \\
&\ -\beta\ \mathbb{D}_{\mathrm{KL}}\!\left[\pi_{\theta}\,\|\,\pi_{\mathrm{ref}}\right]
\end{align}
where:
\begin{itemize}
    \item \( r_i = \frac{\pi_\theta(o_i \mid q)}{\pi_{\theta_{\text{old}}}(o_i \mid q)} \) is the probability ratio, which quantifies how much the new policy \( \pi_\theta \) differs from the old policy \( \pi_{\theta_{\text{old}}} \) for the predicted token \( o_i \). The current policy $\pi_\theta$ is optimized relative to the previous policy $\pi_{\theta_{\text{old}}}$, which serves as the fixed reference from the last optimization step, while a fixed reference policy $\pi_{\text{ref}}$, typically the pretrained model, ensures stability.
    \item \( \mathbb{D}_{\mathrm{KL}}[\cdot] \) denotes the KL divergence penalty, used to regularize the policy updates.
    \item The clipping operation and the minimum function prevent drastic policy changes, thereby ensuring training stability.
\end{itemize}
\vspace{1em}
\phantomsection
\begin{tcolorbox}[
  colback=gray!5!white, 
  colframe=black!75!black, 
  title=\textbf{Prompt used to query the language model},
  sharp corners=south,
  boxrule=0.8pt
]

Answer the question using \textbf{ONLY} the timestamped events.
Your answer must be \textbf{EXACTLY} one of the provided options (A/B/C/D).

\vspace{1em}

\textbf{---} \\
\textbf{[Timestamped Events]} \\
\texttt{<EVENTS>} \\
\textbf{---} \\
\textbf{[Question]} \\
\texttt{<QUESTION>} \\
\textbf{---} \\
\textbf{[Options]} \\
\texttt{<CHOICES>} \\
\textbf{---}  

\vspace{1em}

\textbf{Rules:}
\begin{itemize}
  \item Use \textbf{ONLY} the provided events.
  \item If unsure, choose the option most consistent with the evidence — \textbf{never} output ``N/A'' or ``N''.
  \item Output \textbf{ONLY} the letter of the correct option (A/B/C/D), with \textbf{NO} punctuation, text, or explanations.  
\end{itemize}
\end{tcolorbox}
\label{box:sap}
\vspace{0.5em}
\noindent\textbf{Box~\ref{box:sap}.} This prompt provides the model with strict instructions to select only the letter identifying the correct answer, using the provided timestamped events as support.
\section{Datasets}
\subsection{Audio question answering dataset}
The dataset provided for the challenge \cite{Yang2025} consists of three types of audio/question pairs: Bioacoustics QA \cite{Kim2025}, Temporal Soundscapes QA, and Complex QA \cite{Sakshi2024}. Table \ref{tab:qa_splits} summarizes the proportions of these benchmarks.
\vspace{0.3em}

The first part (Part 1) evaluates how well the models can adapt to diverse acoustic conditions. The second part (Part 2) assesses the model's ability to detect the start, end, and transitions of different acoustic events. The final part (Part 3) evaluates the model’s capability to answer complex questions involving a diverse set of real-world audio scenarios.
\begin{table}[htbp]
\centering
\resizebox{0.7\columnwidth}{!}{%
\begin{tabular}{l|cccc}
\toprule
\textbf{Split} & \textbf{Part1} & \textbf{Part2} & \textbf{Part3} & \textbf{Total} \\
\midrule
train & 740 & 1038 & 6443 & 8221 \\
dev & 224 & 609 & 1633 & 2466 \\
eval & 1480 & 1004 & 2400 & 4884 \\
\bottomrule
\end{tabular}%
}
\caption{Dataset splits for each QA category}
\label{tab:qa_splits}
\end{table}

\subsection{Sound event detection dataset}

The BEATs model have been finetuned on AudioSet strong train set by the authors of \cite{schmid2025effective}.
The logistic regression models are trained and evaluated on subsets of the AudioSet strong evaluation set \cite{asstrong}.
\vspace{0.3em}

Calibration models  are trained on a restricted subset of the AudioSet evaluation set, consisting of 13,596 audio clips (approximately 1,142,064 frames, retaining 84 frames of 40 ms per audio) selected randomly. Furthermore, the evaluation was conducted on another limited subset of only 1,700 audio clips (142,800 frames).

\section{Results}
\label{sec:results}

\subsection{Evaluation metrics}
We optimize our submitted system based on the accuracy, which corresponds to the overall evaluation accuracy (i.e., accuracy computed according to the proportion of the three parts different from the domain average accuracy of the challenge, which is the overall unweighted accuracy).

\subsection{Submitted system}
Our single submission is based on a cascaded model architecture comprising two main modules, totaling 7.7 billion parameters. 
\vspace{0.3em}

First, we use the BEATs audio encoder  \cite{chen2022beats}, finetuned for strong event detection \cite{schmid2025effective} to extract posterior probabilities for each sound event class of the Audioset ontology \cite{gemmeke2017audio} at the segment level. These probabilities are then calibrated before applying median filtering and a threshold of 0.1 to derive event-level predictions. Each prediction is then converted into a text string following the format:
\texttt{\{class\}\_\{start\_time\}\_\{end\_time\}}. These strings are incorporated into a prompt, alongside the corresponding question and answer choices.
\vspace{0.3em}

The resulting prompt is processed by a language model, specifically Qwen2.5-7B-Instruct \cite{qwen2.5}, which was fine-tuned using LoRA (Low-Rank Adaptation) \cite{lora}. In our setup, LoRA injects low-rank trainable matrices into the attention projection layers ($q_{\text{proj}},\ k_{\text{proj}},\ v_{\text{proj}},\ \text{and } o_{\text{proj}}$), allowing us to efficiently adapt the model with minimal parameter updates. We use a moderate rank of $16$ and scaling factor of $32$ to balance adaptation capacity and training stability. We fine-tune our model using GRPO, $|G|=8$, $\epsilon=0.2$, and $\beta=0$. Optimization is performed using a 8-bit quantized AdamW optimizer, with an initial learning rate of $3 \times 10^{-5}$ and a weight decay coefficient of $0.01$. Training is conducted over a single epoch, using a batch size of 8, on a Nvidia A100 GPU with 80 GB of memory.
\vspace{0.3em}

Regarding the inference strategy, we evaluate which answer option makes the complete sequence most probable by estimating the token likelihood, considering only the tokens corresponding to each candidate option.

\subsection{Results on the challenge}
\begin{table}[htbp]
\centering
\resizebox{\columnwidth}{!}{%
\begin{tabular}{l|cccc}
\toprule
\textbf{Model} & \textbf{Part1} & \textbf{Part2} & \textbf{Part3} & \textbf{Total} \\
\midrule
Qwen2-Audio-7B & 30.0 & 39.2 & 49.6 & 45.0 \\
AudioFlamingo2 & 53.9 & 31.7 & 49.5 & 45.7 \\
Gemini-2.0-Flash & 42.0 & 46.3 & 56.6 & 52.5 \\
\midrule
\textbf{Ours} & \textbf{50.4} & \textbf{54.0} & \textbf{67.5} & \textbf{62.6} {\color{moderngreen}\scriptsize (+10.1\%)} \\
\textbf{Ours} (with dev) & 58.5 & 60.0 & 68.3 & 65.4 {\color{moderngreen}\scriptsize (+12.9\%)} \\
\textbf{Ours} (on eval) & \textbf{42.5} & \textbf{50.0} & \textbf{75.3} & \textbf{60.2} {\color{moderngreen}\scriptsize (+3.9\%)} \\
\bottomrule
\end{tabular}%
}
\caption{Comparison of accuracy between the baselines and our method across the three parts and overall, on both the development and evaluation sets. In green is the difference compared to Gemini-2.0-Flash, the best baseline.}
\label{tab:performance_comparison}
\end{table}
As shown in Table \ref{tab:performance_comparison}, our method outperforms the baselines provided by the challenge. The submitted system was trained using the same procedure on both the training and development datasets. Its performance on the development set (a subset of the training set) is reported in the figure, indicated by the label “with dev.”

\subsection{Calibration results}
\label{sec:results_calibration}

Initial evaluation confirm that the BEATs model exhibited poor overall calibration. It tends to be underconfident for certain classes, such as \texttt{Male speech, man speaking} and \texttt{Female speech, woman speaking} and conversely overconfident for others, such as \texttt{Mechanisms}.

\subsubsection{Focus on the example of a specific class : Male speech, man speaking}

\begin{figure}[h]
  \centering
\centerline{\includegraphics[width=0.8\columnwidth]{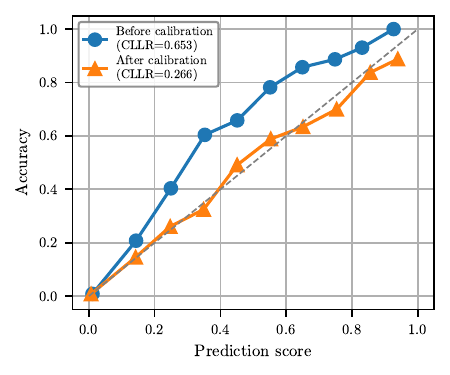}}
  \caption{Reliability curve for the class \texttt{Male Speech, man speaking}.}
  \label{fig:results_calibration}
\end{figure}

In the selected calibration example (Figure \ref{fig:results_calibration}) of the class \texttt{Male Speech, man speaking}, the reliability curve before calibration deviates significantly from the ideal diagonal, indicating poor model calibration. The predicted confidence scores are generally lower than the actual proportion of positive instances, reflecting under-confidence in the model’s predictions. For instance, an average predicted score of 0.6 corresponds in practice to a class occurrence frequency above 0.8.
After calibration, the reliability curve aligns more closely with the diagonal, demonstrating a better match between predicted confidence and observed frequencies. This qualitative improvement is quantitatively supported by a significant decrease in the CLLR score, which drops from 0.653 to 0.266, indicating a notable reduction in calibration error.

\subsubsection{Results on the 447 AudioSet classes}

\begin{figure}[h]
  \centering
  \centerline{\includegraphics[width=0.8\columnwidth]{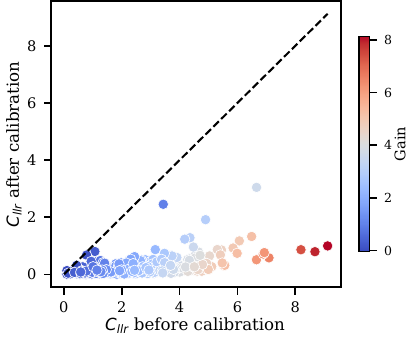}}
  \caption{Class-wise CLLR Comparison on AudioSet – Before vs After Calibration}
  \label{fig:results_calibration_all_classes}
\end{figure}

Figure \ref{fig:results_calibration_all_classes} represents the CLLR before and after calibration for each class in the AudioSet dataset. Each point represents a class, and its color shows the gain achieved through calibration (i.e., the CLLR before calibration minus the CLLR after calibration), with red tones indicating a substantial gain and blue tones indicating a small or negligible one. Most points lie below the diagonal, reflecting a systematic decrease in CLLR after calibration, which is an indication of an overall improvement in the quality of the prediction scores. Some classes that were initially very poorly calibrated (CLLR $>$ 4) see their scores substantially reduced (to below 1), highlighting the ability of the calibration process to correct extreme cases. Moreover, the post-calibration CLLRs are predominantly below 0.5, suggesting good probabilistic adjustment. Finally, the post-calibration scores are much more tightly clustered, indicating a homogenization of performance across classes.
\vspace{0.3em}

These results demonstrate the effectiveness of LLR-based calibration combined with prior adjustment in enhancing both the interpretability and reliability of BEATs model predictions on the AudioSet dataset.
\\
\subsubsection{Impact of calibration on our system}
\label{sec:impact}
The first three lines of Table \ref{tab:calib_table_res} illustrate the impact of threshold selection on system performance. Calibration improves performance, particularly for Part 2, as it allows the same threshold to be effectively applied across different classes. A threshold of 0.1 was selected, as it yielded the highest score on the development set.
\begin{table}[htbp]
\centering
\resizebox{0.5\textwidth}{!}{%
\begin{tabular}{l|c|c|ccc}
\toprule
\textbf{Fine-tuning} & \textbf{Calibration} & \textbf{Threshold} & \textbf{Part 1} & \textbf{Part 2} & \textbf{Part 3} \\
\midrule
NO & NO & 0.05 & 27.2 & 46.5 & 45.1 \\
NO & NO & 0.1  & 31.3 & 46.5 & 44.9 \\
NO & NO & 0.2  & 31.3 & 41.5 & 43.2 \\
NO & YES & 0.05 & 35.3 & 47.9 & 45.0 \\
NO & YES & 0.1  & 35.7 & 51.2 & 44.1 \\
NO & YES & 0.2  & 32.6 & 46.6 & 45.0 \\ 
YES & NO & 0.1 & 51.8 & 49.4 & 64.2 \\
YES & YES & 0.1 & 50.4 & 54.0 & 67.5 \\
\bottomrule
\end{tabular}%
}
\caption{Performance of our system (Qwen2.5-7B-Instruct + BEATS~) on the development set under different training, calibration and threshold settings. Fine-tuning: indicates whether the language model has been fine-tuned.}
\label{tab:calib_table_res}
\end{table}
\vspace{0.3em}

The last two lines compare the benefit of using calibration when the language model is fine-tuned. In Part 1, we observe a slight performance drop with calibration. This can be explained by the fact that the questions in this section primarily involve the classification and detection of marine mammals, which are not among the classes present in AudioSet. However, for the other two parts of the challenge, the questions involve audio samples from datasets whose ontologies are often closely aligned with that of AudioSet. This likely accounts for the performance gains observed when applying calibration.

\subsection{What does GRPO truly encourage ?}
Sometimes the correct answer is in the model’s output, but extracting it in the expected format remains difficult. This raises the question of whether the gain due to fine-tuning with GRPO stems more from enforcing answer formatting than from actual improvements in reasoning about acoustic events.
\vspace{0.2em}

To address this issue, we use the same base model (Qwen2.5-7B-Instruct), fine-tuned with GRPO under three different configurations (Table \ref{tab:comparaison_ev}): (1) events are included neither during training nor inference, (2) events are included only at inference time, and (3) events are included during both training and inference.
\vspace{0.2em}

When acoustic events are not included during training or inference, performance remains high (Table \ref{tab:comparaison_ev}). This suggests that some questions can be answered using the question alone or by eliminating wrong options. Adding events only at inference brings little improvement.
\begin{table}[htbp]
\centering
\resizebox{0.8\columnwidth}{!}{%
\begin{tabular}{c|c|ccc}
\toprule
\textbf{Training} & \textbf{Inference} & \textbf{Part 1} & \textbf{Part 2} & \textbf{Part 3} \\
\midrule
NO & NO & 51.8 & 49.7 & 65.1 \\
NO & YES & 49.1 & 50.9 & 65.5 \\
YES & YES & 50.4 & 54.0 & 67.5 \\
\bottomrule
\end{tabular}%
}
\caption{Comparison of accuracy on the development set after 1 epoch of fine-tuning Qwen2.5-7B-Instruct. Training (respectively, Inference) indicates that events are included in the prompt during training (respectively, inference).}
\label{tab:comparaison_ev}
\end{table}

However, including them during training significantly boosts performance (except for Part 1; see Section \ref{sec:impact}). This shows that GRPO helps the model better use acoustic events to answer questions.
\vspace{0.1em}

The results in the first row show something surprising: even without using audio, a cascaded model performs better than end-to-end models that use audio. This suggests that much of the needed information is already present in the question and answer choices
\vspace{0.1em}

\section{SUMMARY AND FUTURE WORK}
We proposed a large language model for Task 5 of the DCASE 2025 Challenge. The model does not process audio directly but is fine-tuned using GRPO, which allows effective adaptation with limited data, from prompts that include timestamped acoustic events detected by a sound event detection model. We showed that calibrating the likelihood scores for each detected event class improves performance on the challenge.  As future work, we suggest adapting the event priors based on the question being asked or on previously detected events.


\clearpage
\bibliographystyle{IEEEtran}
\bibliography{refs}







\end{document}